\documentclass[aps,pra,twocolumn,showpacs,preprintnumbers,amsmath,amssymb,superscriptaddress,10pt,balancelastpage]{revtex4-1}

\usepackage{amsmath}
\usepackage{bm}% bold math
\usepackage{graphicx}% Include figure files
\graphicspath{{figures/}}
\usepackage{SIunits}
\usepackage{hyphenat}
\usepackage[bookmarks=false]{hyperref}
\usepackage{color}
\usepackage[capitalise]{cleveref}

\definecolor{pink}{RGB}{255,0,255}

\definecolor{ss_color}{rgb}{1,0,0}

\begin{document}

\title{\Large Creation of backdoors in quantum communications via laser damage}

\author{Vadim~Makarov}
\email{makarov@vad1.com}
\affiliation{\mbox{The rest of the authors are listed alphabetically.}}
\affiliation{Institute for Quantum Computing, University of Waterloo, Waterloo, ON, N2L~3G1 Canada}
\affiliation{Department of Physics and Astronomy, University of Waterloo, Waterloo, ON, N2L~3G1 Canada}
\affiliation{\mbox{Department of Electrical and Computer Engineering, University of Waterloo, Waterloo, ON, N2L~3G1 Canada}}

\author{Jean-Philippe~Bourgoin}
\affiliation{Institute for Quantum Computing, University of Waterloo, Waterloo, ON, N2L~3G1 Canada}
\affiliation{Department of Physics and Astronomy, University of Waterloo, Waterloo, ON, N2L~3G1 Canada}

\author{Poompong~Chaiwongkhot}
\affiliation{Institute for Quantum Computing, University of Waterloo, Waterloo, ON, N2L~3G1 Canada}
\affiliation{Department of Physics and Astronomy, University of Waterloo, Waterloo, ON, N2L~3G1 Canada}

\author{Mathieu~Gagn{\' e}}
\affiliation{Department of Engineering Physics and Department of Electrical Engineering, {\' E}cole Polytechnique de Montr{\' e}al, Montr{\' e}al, QC, H3C~3A7 Canada}

\author{Thomas~Jennewein}
\affiliation{Institute for Quantum Computing, University of Waterloo, Waterloo, ON, N2L~3G1 Canada}
\affiliation{Department of Physics and Astronomy, University of Waterloo, Waterloo, ON, N2L~3G1 Canada}
\affiliation{Quantum Information Science Program, Canadian Institute for Advanced Research, Toronto, ON, M5G~1Z8 Canada}

\author{Sarah~Kaiser}
\affiliation{Institute for Quantum Computing, University of Waterloo, Waterloo, ON, N2L~3G1 Canada}
\affiliation{Department of Physics and Astronomy, University of Waterloo, Waterloo, ON, N2L~3G1 Canada}

\author{Raman~Kashyap}
\affiliation{Department of Engineering Physics and Department of Electrical Engineering, {\' E}cole Polytechnique de Montr{\' e}al, Montr{\' e}al, QC, H3C~3A7 Canada}

\author{Matthieu~Legr{\' e}}
\affiliation{ID~Quantique SA, Chemin de la Marbrerie~3, 1227 Carouge, Geneva, Switzerland}

\author{Carter~Minshull}
\affiliation{Institute for Quantum Computing, University of Waterloo, Waterloo, ON, N2L~3G1 Canada}

\author{Shihan~Sajeed}
\affiliation{Institute for Quantum Computing, University of Waterloo, Waterloo, ON, N2L~3G1 Canada}
\affiliation{\mbox{Department of Electrical and Computer Engineering, University of Waterloo, Waterloo, ON, N2L~3G1 Canada}}

\date{September 18, 2016}

\begin{abstract}
Practical quantum communication (QC) protocols are assumed to be secure provided implemented devices are properly characterized and all known side channels are closed. We show that this is not always true. We demonstrate a laser-damage attack capable of modifying device behaviour on-demand. We test it on two practical QC systems for key distribution and coin-tossing, and show that newly created deviations lead to side channels. This reveals that laser damage is a potential security risk to existing QC systems, and necessitates their testing to guarantee security.
\end{abstract}

\maketitle

\noindent Cryptography, an art of secure communication, has traditionally relied on either algorithmic or computational complexity \cite{singh1999}. Even the most state-of-the-art classical cryptographic schemes do not have a strict mathematical proof to ascertain their security. With the advance of quantum computing, it may be a matter of time before the security of the most widely used public-key cryptography protocols is broken \cite{shor1997}. Quantum communication (QC) protocols, on the other hand, have theoretical proofs of being unconditionally secure \cite{bennett1984,lo1999,gottesman2004,barz2012,collins2014,lunghi2013,grice2015}. In theory, their security is based on the assumption of modeled behaviour of implemented equipment. In practice, the actual behaviour often deviates from the modeled one, leading to a compromise of security as has been seen so far in case of quantum key distribution (QKD) \cite{bennett1992b,makarov2006,qi2007,lydersen2010a,xu2010,sun2011,sajeed2015a}. However, it is widely assumed that as long as these deviations are properly characterized and security proofs are updated accordingly \cite{gottesman2004,fung2009}, implementations are unconditionally secure. In this work we show that satisfying this during the initial installation only is not enough to guarantee security. Even if a system is perfectly characterized and deviations are included into the security proofs, an adversary can still create a new deviation on-demand and make the system insecure.

\begin{figure*}
  \includegraphics{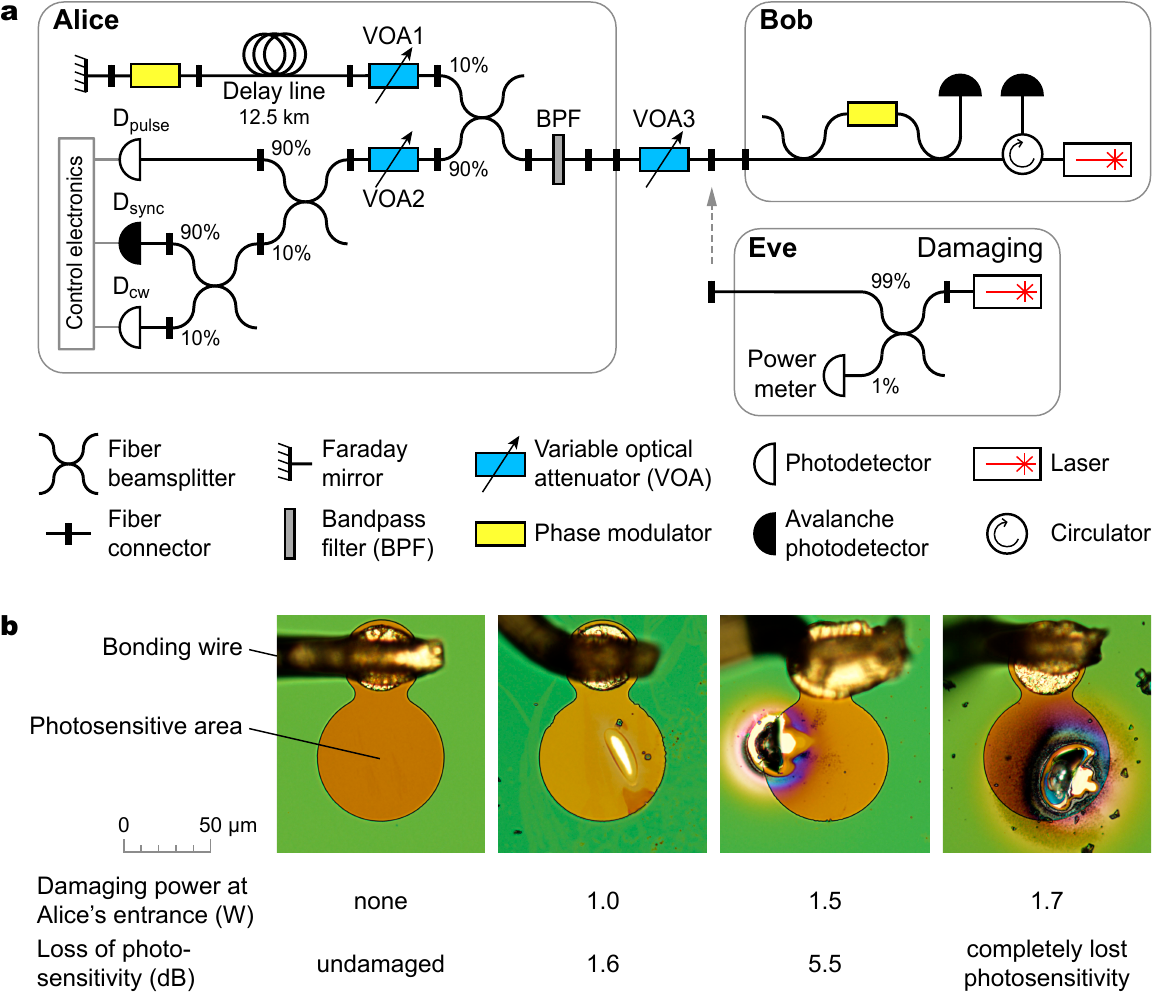}
  \caption{{\bf Attack on fiber-optic system Clavis2.} {\bf a},~Experimental setup. The system consists of Alice and Bob connected by a lossy fiber communication channel (simulated by variable optical attenuator VOA3). Bob sends to Alice pairs of bright coherent optical pulses, produced by his laser and two fiber arms of unequal length \cite{stucki2002,idqclavis2specs}. Alice uses fiber beamsplitters to divert parts of incoming pulse energy to monitoring detector D$_\text{pulse}$, synchronization detector D$_\text{sync}$ and line-loss measurement detector D$_\text{cw}$. She prepares quantum states by phase-modulating the pulses, reflecting them at a Faraday mirror and attenuating to single-photon level with VOA1. Bob measures the quantum states by applying his basis choice via phase modulator and detecting outcome of quantum interference with single-photon avalanche photodetectors. Eve's damaging laser is connected to the channel manually. BPF, bandpass filter. {\bf b},~Pulse-energy-monitoring photodiode before and after damage. Brightfield microphotographs show top-view of decapsulated photodiode chips. The last two samples have holes melted through their photosensitive area. Scattered dark specks are debris from decapsulation.}
  \label{fig:fiber-optic-setup}
\end{figure*}

Before going into details on how the adversary may do it, let's consider a few examples of deviations and their consequences. For example, a calibrated optical attenuator is required to set a precise value of the outgoing mean photon number $\mu$ in the implementations of ordinary QKD \cite{stucki2002,idqclavis2specs}, decoy-state QKD \cite{bourgoin2015}, coherent-one-way QKD \cite{walenta2014}, measurement-device-independent QKD \cite{tang2014}, continuous-variable QKD \cite{jouguet2013a}, digital signature \cite{collins2014}, relativistic bit commitment \cite{lunghi2013}, coin-tossing \cite{pappa2014} and secret-sharing \cite{grice2015} protocols. An unexpected increase of this optical component's attenuation may cause a denial-of-service. However, a reduction in attenuation will increase $\mu$, leading to a compromise of security via attacks that rely on measurement of multi-photon pulses \cite{felix2001,sajeed2015}. E.g.,\ in QKD and secret-sharing this will allow eavesdropping of the key, and in bit commitment cheating the committed bit value. Some implementations use a detector for time synchronization \cite{stucki2002,idqclavis2specs,walenta2014,tang2014,jouguet2013a,lunghi2013,pappa2014,grice2015}. Desensitizing it may result in the denial-of-service. However, several implementations require a calibrated monitoring detector for security purposes \cite{stucki2002,idqclavis2specs,walenta2014,jouguet2013a,lunghi2013,pappa2014,grice2015}. A reduction in its sensitivity may lead to security vulnerabilities such as a Trojan-horse attack that reads the state preparation \cite{vakhitov2001}. This leaks the key in QKD, increases the cheating probability in coin-tossing \cite{sajeed2015}, leaks the program and client's data in quantum cloud computing \cite{barz2012} and allows forging of digital signatures \cite{collins2014}. Many implementations use beamsplitters and rely on their pre-characterized splitting ratio (e.g.,\ \cite{stucki2002,idqclavis2specs,bourgoin2015,walenta2014,jouguet2013a,lunghi2013,pappa2014}). A shift in the splitting ratio may lead to either the denial-of-service or security vulnerabilities (e.g.,\ \cite{li2011a} or one of the above-mentioned attacks). A shift in characteristics of a phase modulator or a Faraday mirror may create imperfect qubits that will result in the denial-of-service or a breach in security \cite{xu2015,xu2010,sun2011}. If the dark count rate of single-photon detectors is increased, it may lead to the denial-of-service \cite{bugge2014}. Even in device-independent QKD (DI-QKD) \cite{acin2006}, the absence of information-leakage channels and memory is assumed \cite{barrett2013}. Thus, there is a risk these assumptions may be compromised by deviations in device characteristics. To give a speculative illustration, let's suppose detectors in DI-QKD emit light on detection \cite{kurtsiefer2001,pinheiro2015,meda2016}, and to prevent this leaking information about detection results, spectral filters and optical isolators are added to the devices. Then, unexpected deviations in characteristics of the latter components become important for security. In summary, quantum communication systems rely on multiple characteristics of many components for their correct operation, and a deviation might lead to severe security consequences.

\begin{figure*}
  \includegraphics{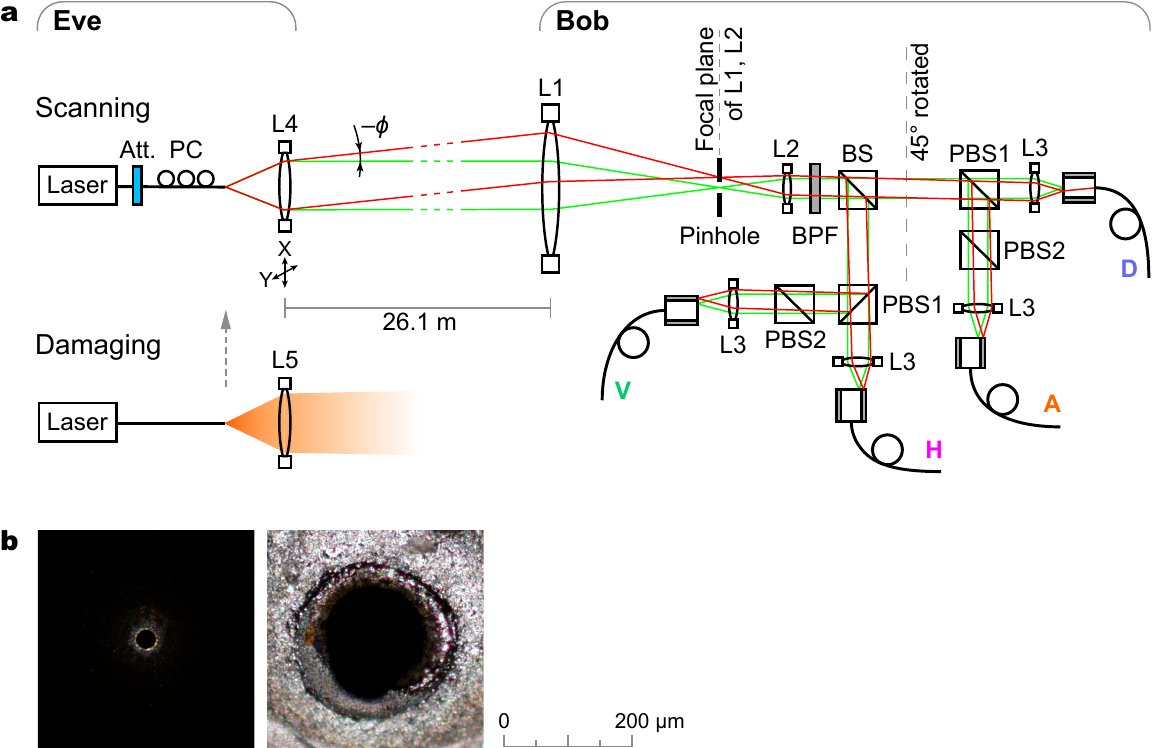}
  \caption{{\bf Attack on free-space QKD system.} {\bf a},~Experimental setup. QKD receiver Bob consists of two lenses L1, L2 reducing input beam diameter, 50:50 beamsplitter BS, and two arms measuring photons in HV and DA polarizations using polarizing beamsplitters PBS \cite{sajeed2015a,bourgoin2015}. Photons are focused by lenses L3 into multimode fibers leading to single-photon detectors. Setup drawing is not to scale. Eve's apparatus contains a scanning laser source that tilts the beam angle $(\phi, \theta)$ by laterally shifting lens L4. Green marginal rays denote initial Eve's alignment, replicating the alignment Alice--Bob at $\phi = \theta = 0$. Red marginal rays show a tilted scanning beam missing fiber cores V, H, A, but coupling into D. Eve's damaging laser source can be manually inserted in place of the scanning source. Att.,\ attenuator; PC, polarization controller. {\bf b},~Spatial filter before and after damage. Darkfield microphotographs show front view of the pinhole. See \cref{sec:video-pinhole-damage} for real-time video recording of laser damage to the pinhole inside Bob.}
  \label{fig:free-space-setup}
\end{figure*}

In classical communications, there is no real concern about the possibility of a shift in device characteristics. Classical devices' security-critical parts can be physically separated from the communication channel and isolated from physical access by the adversary \cite{tempest-2-95}. However the front-end of a quantum communication system is essentially an analog optical system connected to the channel (at least, at our present level of the technology), and is easily accessible by the adversary. The latter can shoot a high-power laser from the communication channel to alter system component characteristics via laser damage \cite{bugge2014}. The question is what will this achieve? Will the adversary break some component needed for operation and cause the denial-of-service (which is not a useful outcome for her), or will she change some component in such a way as to facilitate a compromise of security? Further, will the security compromise be only possible in theory or be practical with today's technology? This cannot be predicted in advance, because system implementations contain many components and their laser damage thresholds and failure behavior are generally not precisely known. To assess the risk for quantum communications, we have performed tests on two extensively characterized, completely different and widely used implementations: a commercial fiber-optic system for QKD and coin-tossing with phase-encoded qubits \cite{stucki2002,idqclavis2specs}, and a free-space system for QKD with polarization-encoded qubits \cite{bourgoin2015}. In both systems, we have unfortunately observed the best possible outcome for the adversary. After the laser damage, the systems' security has become compromisable with today's technology.

% Fiber-optic experiment
\medskip
{\noindent \bf Laser damage in fiber-optic system.} As a representative of a fiber-optic quantum communication implementation, we chose a plug-and-play QKD \cite{stucki2002} and loss-tolerant quantum coin tossing (QCT) \cite{pappa2014}. Both were implemented using a commercial system Clavis2 from ID~Quantique \cite{idqclavis2specs}. In both cases, Bob sends bright light pulses to Alice. Alice randomly encodes her secret bits by applying one out of four phases ($0, \frac{\pi}{2}, \pi, \frac{3\pi}{2}$), attenuates the pulses and reflects them back to Bob (\cref{fig:fiber-optic-setup}a). The security of both protocols requires an upper bound on the mean photon number $\mu$ coming out of Alice. Otherwise, an eavesdropper Eve can perform a Trojan-horse attack \cite{vakhitov2001} by superimposing extra light to the bright pulses on their way to Alice from Bob. If Alice is unaware of this and applies the same attenuation, then light coming out of her has a higher $\mu$ than allowed by the security proofs \cite{gottesman2004}, making the implementations insecure. It is thus crucial for the security of both protocols that Alice monitors the incoming pulse energy. This is achieved by employing a pulse-energy-monitoring detector (D$_\text{pulse}$ in \cref{fig:fiber-optic-setup}a). A portion of the incoming light is fed to D$_\text{pulse}$ such that whenever extra energy is injected, an alarm is produced \cite{sajeed2015}. The sensitivity of D$_\text{pulse}$ is factory-calibrated, thus closing the side-channel associated with the Trojan-horse attack. 

Our testing showed that this countermeasure is vulnerable to laser damage. During normal QKD operation, we disconnected the fiber channel Alice--Bob temporarily and connected Eve (\cref{fig:fiber-optic-setup}a). She then injected $1550~\nano\meter$ laser light from an erbium-doped fiber amplifier for $20$--$30~\second$, delivering continuous-wave (c.w.)\ high power into Alice's entrance. $44\%$ of this power reached the fiber-pigtailed InGaAs \mbox{p-i-n} photodiode D$_\text{pulse}$ (JDSU EPM 605LL), and damaged it partially or fully. It became either less sensitive to incoming light (by $1$--$6~\deci\bel$ after $0.5$--$1.5~\watt$ illumination at Alice's entrance) or completely insensitive (after $\geq 1.7~\watt$). The physical damage is shown in \cref{fig:fiber-optic-setup}b. No other optical component was damaged at this power level. We repeated the experiment with 6 photodiode samples. In half of these trials, QKD continued uninterrupted and kept producing more key after we reconnected the channel back to Bob, as if nothing has happened. In the other half, a manual software restart was needed. However, in all the trials the damage was sufficient to permanently open the system up to the Trojan-horse attack. As modeled in Ref.~\onlinecite{sajeed2015}, in the QKD protocol, Eve can eavesdrop partial or full key using today's best technology if the sensitivity of D$_\text{pulse}$ drops by more than $5.6~\deci\bel$. In the QCT implementation, a sensitivity reduction by $2.6~\deci\bel$ can increase Bob's cheating probability above a classical level, removing any quantum advantage of coin-tossing. Laser damage thus compromises both the QKD and QCT implementations. See \cref{sec:experiment-details-fiber-optic} for details. 

\begin{figure*}
  \includegraphics{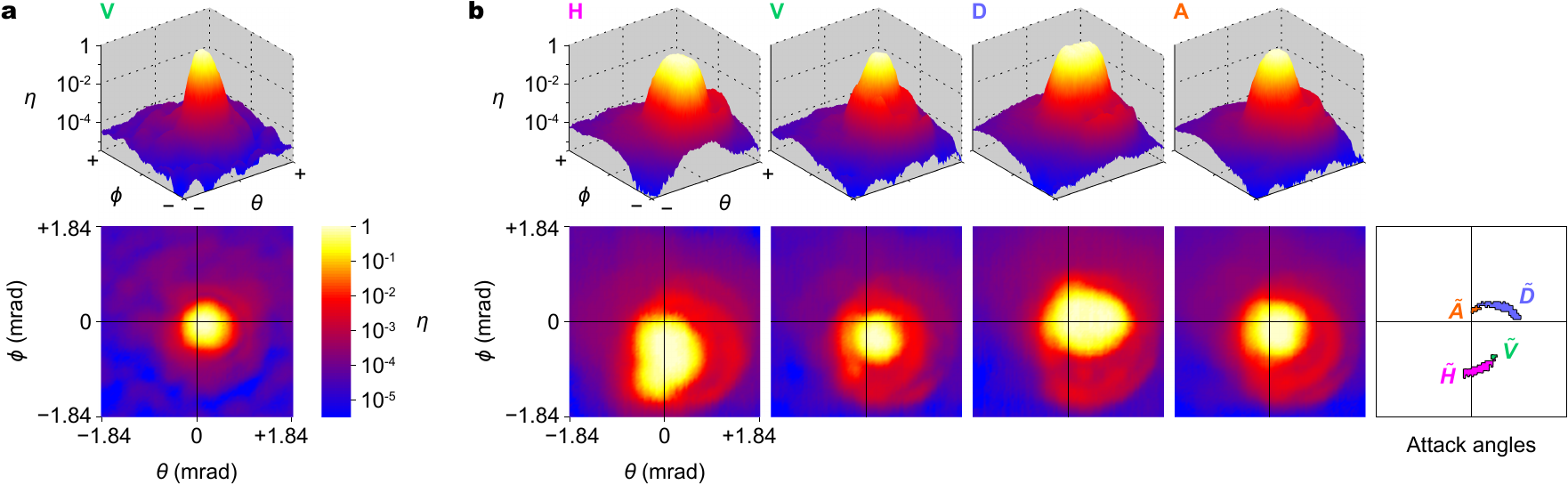}
  \caption{{\bf Efficiency-mismatch side-channel opened after laser damage in free-space QKD system}. Each pair of 3D--2D plots shows normalised photon detection efficiency $\eta$ in a receiver channel versus illuminating beam angles $\phi$ and $\theta$. {\bf a},~Before laser damage, the angular dependence is essentially identical between the four channels \cite{sajeed2015a}. Plot for one channel (V) before damage is shown. {\bf b},~After the laser damage, the four receiver channels H, V, D, A exhibit unequal sensitivity to photons outside the middle area around $\phi = \theta = 0$. The last plot shows angular ranges for targeting the four detectors that satisfy conditions for the faked-state attack.}
  \label{fig:free-space-mismatch-plots}
\end{figure*}

% Free-space experiment
\medskip
{\noindent \bf Laser damage in free-space system.} As a representative of free-space quantum communication, we chose a long-distance satellite QKD prototype operating at $532~\nano\meter$ wavelength \cite{bourgoin2015} employing Bennett-Brassard 1984 (BB84) protocol \cite{bennett1984}. At each time slot, Alice randomly sends one out of four polarizations: horizontal (H), vertical (V), $+45\degree$ (D), or $-45\degree$ (A) using a phase-randomized attenuated laser. Bob randomly measures in either horizontal-vertical (HV) or diagonal-antidiagonal (DA) basis, using a polarization-beamsplitter receiver (\cref{fig:free-space-setup}a). It has been shown in Ref.~\onlinecite{sajeed2015a} that an eavesdropper can, in practice, tilt the beam going towards Bob by an angle $(\phi, \theta)$ such that the beam misses, partially or fully, the cores of fibers leading to three detectors while being relatively well coupled into the core leading to the fourth detector, as illustrated in \cref{fig:free-space-setup}a. This happens because real-world optical alignments are inherently imperfect and manufacturing precision is finite. By sending light at different spatial angles, the eavesdropper can have control over Bob's basis and measurement outcome and steal the key unnoticed \cite{makarov2005,lydersen2010a,sajeed2015a}. This attack can be prevented by placing a spatial filter or `pinhole' at the focal plane of lenses L1 and L2, as shown in \cref{fig:free-space-setup}a \cite{sajeed2015a}. Since the pinhole limits the field of view, any light entering at a higher spatial angle is blocked and Eve no longer has access to the target angles required to have control over Bob. As was demonstrated in Ref.~\onlinecite{sajeed2015a}, a pinhole of $25~\micro\meter$ diameter eliminates this side-channel by making the angular efficiency dependence identical between the four detectors (\cref{fig:free-space-mismatch-plots}a).

Our testing showed that this countermeasure is destroyed by laser damage. From a distance of $26.1~\meter$, we shot an $810~\nano\meter$ collimated laser beam delivering a $10~\second$ pulse of $3.6~\watt$ c.w.\ power at the pinhole inside Bob's setup. The intensity there was sufficient to melt the material ($13~\micro\meter$ thick stainless steel) and enlarge the hole diameter to $\approx\! 150~\micro\meter$. The state of the pinhole before and after damage is shown in \cref{fig:free-space-setup}b, and a real-time video of the damage process is shown in \cref{sec:video-pinhole-damage}. Although Bob was up and running in photon counting mode during the test, none of his other components were damaged. With this larger pinhole opening, it was again possible to send light at angles that had relatively higher mismatches in efficiency, as shown in \cref{fig:free-space-mismatch-plots}b. This enabled a faked-state attack under realistic conditions of channel loss in $1$--$15~\deci\bel$ range with quantum bit error ratio (QBER) $< 6.6\%$. Thus laser damage completely neutralizes this countermeasure, and makes this free-space QKD system insecure. See \cref{sec:experiment-details-free-space} for details.

% Discussion and conclusion
\medskip
{\noindent \bf Discussion.} The crucial step of the attack, creating the deviation in device characteristics, has thus been experimentally demonstrated for both systems tested. We repeated this step several times and confirmed that laser power above a certain value ($1.7~\watt$ in fiber-optic system and $3.6~\watt$ in free-space one) always destroys the security-critical component, without inflicting any collateral damage that could result in the denial-of-service. After this, building a complete eavesdropper would be a realistic if time-consuming task \cite{gerhardt2011}.

In our testing, we haven't done anything that Eve could not do in the real world. She could buy a copy of each system, rehearse her attacks, then attack an installed system of the same model. By Kerckhoffs' principle \cite{kerckhoffs1883}, Eve is assumed to know the system characteristics and results of damage precisely. In practice when attacking installed devices, if she needs to measure their characteristics, she may probe them remotely by imaging, reflectometry \cite{vakhitov2001}, and watching public communication Alice--Bob \cite{makarov2005,gerhardt2011}.

At present, no quantum communication system has countermeasures specifically designed to stop laser-damage attack, neither do they have a mechanism to check all possible deviations in device characteristics from the modeled values. Countermeasures to other attacks do not prevent this attack, in fact they become weak points as our experimental study shows. Development of necessary countermeasures is complicated by the fact that Eve can use a laser with different characteristics: power, timing (e.g., short-pulsed laser induces different damage mechanisms than c.w.\ thermal damage we have observed \cite{wood2003}), wavelength, polarization. Eve can attack the systems in different phases of their operation including powered-off state, which can control what component is damaged. We have experimentally observed dependence of damage on the laser timing profile, as detailed in \cref{sec:damage-control-fiber-optic}, where we show that some profiles have resulted in the denial-of-service but some in a successful attack. We stress that Eve will select the illumination regime that results in the successful attack, if such regime exists at all. Any countermeasure must thus be tested in all possible illumination regimes. Possible directions of development include a passive optical power limiter \cite{tutt1993.ProgQuantElectr-17-299}, a single-use `fuse' that permanently breaks the optical connection if a certain power is exceeded, battery-powered active monitoring supplemented with wavelength filtering, or an optical isolator (for Alice that uses one-way light propagation \cite{bourgoin2015,walenta2014,tang2014,jouguet2013a,collins2014,barz2012,acin2006,gerhardt2011}). Hardware self-characterization may be promising \cite{lydersen2011a}, however to protect from an arbitrary damage it must monitor a potentially large number of hardware parameters.

It is an interesting question if risk for untested system designs can be estimated. As we have discussed, any given system design contains many optical components with unknown damage characteristics. The outcome of damage (denial-of-service or successful attack) is thus impossible to predict prior to testing. Then, if some of the system designs chosen at random were tested, the risk for the remaining untested designs could be calculated by Bayesian statistics \cite{gelman2004}. Unfortunately, truly random choice is impractical to implement with the current state of quantum communications research and limited sample availability. We have instead tested the two system designs that were available in our lab. This biased the choice towards more mature and older designs. Although this unknown bias makes the Bayesian analysis inapplicable, we find it illustrative to consider the risk figure that would have applied if the choice were random. With zero systems tested, the Bayesian probability that at least $20\%$ of the untested system designs (assuming at least 50 of them exist) are vulnerable to this attack is $70.4\%$ ($80\%$), assuming a Jeffreys (uniform) prior. If two randomly chosen system designs were tested with two positive outcomes, this probability would have increased greatly to $98.9\%$ ($98.6\%$). Note that the security risk is generally high, which is in stark contrast with the very low expected theoretical risk \cite{lo1999,gottesman2004,fung2009}.

We have experimentally demonstrated laser damage as a new eavesdropping tool that alters parameters of a well-characterized quantum communication system. Any modification of system characteristics might compromise the security either directly by leading to an attack as we have demonstrated, or indirectly by shifting some parameter in the security proof so it would no longer apply. Existing security proofs do not accommodate this, neither do existing systems have any countermeasure implemented against this. Our results thus reveal the potential security risk for other existing systems, which should be tested against this attack.

\bigskip

{\bf \noindent Acknowledgements}. We thank Q.~Liu, E.~Anisimova and O.~Di~Matteo for early experimental efforts, S.~Todoroki, N.~L{\" u}tkenhaus, M.~Mosca, Y.~Zhang, L.~Lydersen and S.~Lydersen for discussions. This work was supported by the US Office of Naval Research, Industry Canada, CFI, Ontario MRI, NSERC, Canadian Space Agency, ID~Quantique, European Commission's FET QICT SIQS project, EMPIR 14IND05 MIQC2 project, and CryptoWorks21. We acknowledge using University of Waterloo's Quantum NanoFab. P.C.\ was supported from Thai DPST scholarship. J.-P.B.\ was supported by FED~DEV.

{\bf \noindent Conflicts of interest.} A part of this study was supported and M.L.\ was employed by ID~Quantique. The company has been informed prior to this publication, and is developing countermeasures for their affected QKD system. The other authors declare no competing financial interests.

{\bf \noindent Author contributions.} V.M.\ conceived and led the study. S.K.\ implemented the fiber-optic experiment. S.S.\ implemented the free-space experiment and contributed to the fiber-optic experiment. P.C.\ contributed to the free-space experiment. M.G.\ contributed to the fiber-optic experiment. C.M.\ made minor contributions to the free-space experiment. M.L.\ provided and supported the fiber-optic QKD system under test. T.J.\ and J.-P.B.\ provided the free-space QKD receiver under test and contributed to the free-space experiment. R.K.\ provided the fiber laser facility and co-supervised the fiber-optic experiment. S.S.\ and V.M.\ wrote the article, with contributions from all authors.

\appendix

\section{Laser-damage experiment on fiber-optic system}
\label{sec:experiment-details-fiber-optic}

\noindent In our experiment, we damaged D$_\text{pulse}$ during QKD operation, trying not to interrupt it. The system was allowed to start up and produce a secret key for several QKD cycles, using BB84 protocol \cite{bennett1984}. To perform laser damage, we disconnected the channel for $2$--$3~\minute$, giving us enough time to apply high power to Alice, and then reconnected the channel. We tried this at different points in the QKD operation cycle. Sometimes the software recovered and resumed QKD, and sometimes it got stuck in recalibration routines. In the latter case, a manual software restart resumed QKD. Owing to a limited number of trials, we did not perfect this timing aspect.

We tested a total of 6 photodiode samples. We damaged each of them by applying high power laser light at Alice's entrance. We then used the manufacturer's factory-calibration software to measure how much extra signal power (compared to the pre-calibrated power level) could be injected without triggering the alarm \cite{sajeed2015}. This quantified the reduction in sensitivity due to the damage. Three samples were exposed twice to a progressively higher power. For example, one sample was first exposed to $0.5~\watt$ power at Alice's entrance that reduced its photosensitivity by $1~\deci\bel$, then to $0.75~\watt$ power that reduced its photosensitivity by $6~\deci\bel$. For the other two samples these numbers were $0.75~\watt$ with no change in sensitivity then $1.0~\watt$, $1.6~\deci\bel$ (shown in 2nd microphotograph in \cref{fig:fiber-optic-setup}b); $1.0~\watt$, $5~\deci\bel$ then $1.5~\watt$, $5.5~\deci\bel$ (shown in 3rd microphotograph in \cref{fig:fiber-optic-setup}b). For the remaining three samples, $1.7~\watt$ was applied at Alice's entrance, and D$_\text{pulse}$ completely lost photosensitivity, becoming electrically either a large resistor (shown in 4th microphotograph in \cref{fig:fiber-optic-setup}b) or an open circuit. After we were done with each sample, we used the same manufacturer's factory-calibration software to pre-calibrate the sensitivity of the next undamaged D$_\text{pulse}$ sample, following the factory procedure.

No other component in Alice was damaged during these trials. We also tested some components separately. FC/PC and FC/APC optical connectors used in Alice and in the channel withstood $3~\watt$ c.w., while copies of Alice's 10:90 fiber beamsplitters (AFW Technologies FOSC-1-15-10-L-1-S-2) withstood up to $8~\watt$ c.w.\ with no damage.

\begin{figure}
  \includegraphics{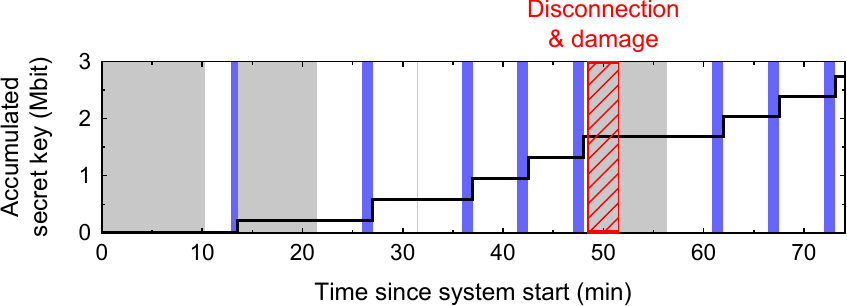}
  \caption{{\bf Fiber-optic QKD system operation during laser damage.} The plot shows accumulated secret key amount versus time. Grey bands denote the system performing recalibration routines, white bands denote the quantum bit sending and receiving, and blue (darker) bands denote classical post-processing. All this information was extracted from the QKD system log files after the experiment. The band hatched in red denotes the time when the fiber channel Alice--Bob was temporarily disconnected and the laser damage to Alice was done by $1.7~\watt$ laser power, resulting in D$_\text{pulse}$ becoming an open circuit with no photosensitivity.}
  \label{fig:fiber-optic-accumulated-key}
\end{figure}

\Cref{fig:fiber-optic-accumulated-key} summarizes a system operation log when it recovered automatically after the damage that made the photodiode an open-circuit with no photosensitivity. In the current system implementation, this represents an ideal outcome for an attacker.

For damaging and component tests, Eve used an erbium-doped fiber amplifier seeded from a $1550.7~\nano\meter$ laser source (EDFA; IPG Photonics ELR-70-1550-LP). She injected $0$--$2~\watt$ c.w.\ power at Alice's entrance. The injected power was monitored with a 1:99 fiber beamsplitter tap and a power meter (\cref{fig:fiber-optic-setup}a). A manually operated shutter at the output of EDFA allowed to ramp the power up and down smoothly between 0 and the target level, with tens of milliseconds transition time. The spectral characteristics of EDFA's built-in seed laser did not precisely match the passband of the BPF at Alice's entrance ($1551.32$--$1552.12~\nano\meter$ passband at $-0.5~\deci\bel$ level, $< 0.7~\deci\bel$ insertion loss; AFW Technologies BPF-1551.72-2-B-1-1). We therefore removed the BPF for the duration of experiment. The BPF was separately tested in-passband using a different EDFA (PriTel LNHPFA-37) with a narrowband seed laser, and passed more than $1~\watt$ c.w.\ with no damage.

The system QKD software (`QKD Sequence' application \cite{idqclavis2specs}) set the variable attenuator VOA2 at $2~\deci\bel$. Thus, $44\%$ of Alice's incoming light impinged D$_\text{pulse}$, while smaller fractions impinged D$_\text{sync}$ and D$_\text{cw}$. The alarm threshold of D$_\text{pulse}$ is calibrated when the system is assembled at the factory, and is not changed after that \cite{sajeed2015}. VOA3 introduced channel loss of $1.87~\deci\bel$, to simulate the effect of $\approx\! 9~\kilo\meter$ long fiber line Alice--Bob.

The QKD system Clavis2 normally operates automatically in cycles consisting of sending and receiving quantum states until either the memory buffer is full or photon detection efficiency has dropped significantly. It then uses the classical link Alice--Bob to post-process the detected data and distill the secret key \cite{bennett1992b}. Each cycle takes several minutes. If the last QKD cycle was interrupted because the detection efficiency was too low, or the key distillation failed, the system returns to start-up routines such as timing recalibration \cite{jain2011} before it resumes sending quantum states. This happens often in normal operation, because of naturally occurring drift of hardware and channel parameters. The software generally tries to recover automatically from various error conditions, to provide long-term unattended operation \cite{stucki2011}.

\medskip

{\bf \noindent Predicted attacks on fiber-optic system with damaged pulse-energy-monitoring photodiode.} As modeled in Ref.~\onlinecite{sajeed2015}, for BB84 QKD protocol Eve can eavesdrop partial or full key information using today's best photonics technologies when the sensitivity of D$_\text{pulse}$ has dropped by $4.3$--$5.6~\deci\bel$, given that communication channel loss Alice--Bob is in a $1$--$7~\deci\bel$ range. (This corresponds to a multiplication factor $x$ in the range of $2.7$--$3.6$, see Fig.~11 in Ref.~\onlinecite{sajeed2015}.) If we assume that Eve's equipment is only limited by the laws of quantum mechanics, then she can extract the full key information after only $0.4$--$0.8~\deci\bel$ reduction in sensitivity ($x$ of $1.1$--$1.2$). Similarly, for QCT with a dishonest Bob only limited by the quantum mechanics, all the quantum advantages of the protocol are eliminated if sensitivity reduction of $2.6~\deci\bel$ is obtained in Alice ($x = 1.805$), for a $15~\kilo\meter$ long communication channel. For a $10~\deci\bel$ sensitivity reduction, Bob's cheating probability approaches unity \cite{sajeed2015}. Since we have surpassed the above sensitivity reduction thresholds in our laser damage experiment, we consider the security of both QKD and QCT implementation compromised.

\section{Laser-damage experiment on free-space QKD system}
\label{sec:experiment-details-free-space}

\noindent In order to neutralize the effect of the pinhole and reproduce the side-channel of spatial-mode detector-efficiency mismatch, our experiment consisted of three steps. Firstly, we performed scanning to certify that the system is secure against this side-channel. Secondly, we laser-damaged the pinhole to open the side-channel. Finally, we performed scanning again to demonstrate that the system's security has been compromised. In all three steps, Eve was placed at a distance of $26.1~\meter$ away from Bob and the steps were performed in sequence without making any interactions with Bob.

The first step involved changing the outgoing beam's angle $(\phi, \theta)$ emitted from Eve's scanning setup shown in \cref{fig:free-space-setup}a, then recording the corresponding count rate at all four detectors in Bob. This step is identical to that in Ref.~\onlinecite{sajeed2015a}. The scanning result is shown in \cref{fig:free-space-mismatch-plots}a, where a pair of 3D--2D plots shows the normalized photon detection efficiency in one receiver channel versus the illuminating beam angles $\phi$ and $\theta$. With the pinhole in place, the angular dependence of efficiency is essentially identical between the four channels, hence only a plot for channel V is shown. No measurable amount of efficiency mismatch was found and no attack angles existed \cite{sajeed2015a}.

Then as the second step, Eve's scanning setup was replaced with the damaging setup. The latter contained a $810~\nano\meter$ laser diode (Jenoptik JOLD-30-FC-12) pumped by a current-stabilized power supply and connected to $200~\micro\meter$ core diameter multimode fiber. It provided continuously adjustable $0$ to $30~\watt$ c.w.\ power into the fiber. An almost-collimated free-space beam was subsequently formed by a plano-convex lens L5 (Thorlabs LA1131-B; \cref{fig:free-space-setup}a). The beam's intensity was nearly uniformly distributed across Bob's L1 ($50~\milli\meter$ diameter achromatic doublet, Thorlabs AC508-250-A), with less than $\pm 10\%$ intensity fluctuation across Bob's input aperture. Transmission of L1 was about $82\%$, owing to its antireflection coating being designed for a different wavelength band. In the test detailed here, the power delivered at the pinhole plane was $3.6~\watt$, sufficient to reliably produce a hole of $\approx\! 150~\micro\meter$ diameter in less than $10~\second$ in a standard stainless-steel foil pinhole (Thorlabs P25S). We tested several pinholes and found that this power always made the hole. We also tested that power decreased to $2.0~\watt$ still produced a hole. No other component in Bob was damaged during the tests. Bob's lenses L3 received $\sim\! 1~\micro\watt$ power each, and single-photon detectors only received on the order of a few $\nano\watt$ each, mainly owing to the presence of BPF after the pinhole. The BPF was used by Bob to increase the signal-to-noise ratio during QKD by heavily attenuating all light outside the $531$--$533~\nano\meter$ passband (it consisted of two stacked filters, Thorlabs FESH0700 followed by Semrock LL01-532-12-5) \cite{bourgoin2015}. While the damaging beam was on, the detectors counted at their saturation rate of $\sim\! 35~\mega\hertz$, which did not look abnormal to Bob as this sometimes occurs naturally owing to atmospheric conditions (during sunset, sunrise, fog). We remark that this type of detector usually survives tens of $\milli\watt$ for a short time \cite{sauge2011,bugge2014}. Even if we had to use a wavelength within the BPF's passband, detector exposure to higher power could likely be avoided by shaping Eve's damaging beam.

After the damage, as the third step we replaced the damaging setup with the scanning setup again, and performed the final scanning of Bob's receiver with the damaged pinhole. The results are shown in \cref{fig:free-space-mismatch-plots}b. Now, the four receiver channels H, V, D, A exhibited unequal sensitivity to photons outside the middle area around $\phi = \theta = 0$. These efficiency plots were different from those measured in Ref.~\onlinecite{sajeed2015a} without the pinhole, because of extra scattering at the edges of our laser-enlarged pinhole.

\medskip

\begin{figure}
  \includegraphics{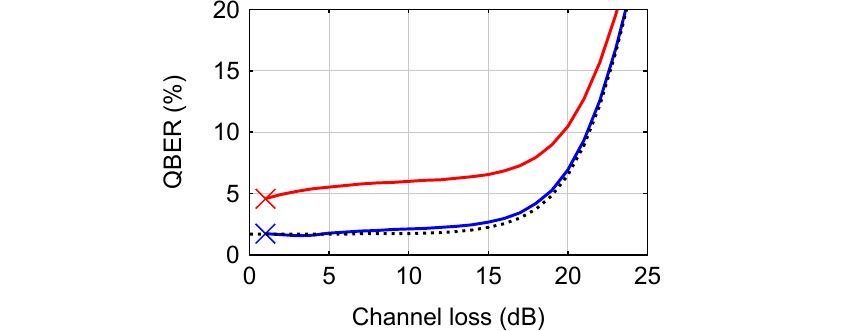}
  \caption{{\bf Modeled QBER observed by Bob in free-space QKD system.} The dotted curve shows QBER without Eve. At lower channel loss, the QBER is due to imperfect fidelity, while at higher channel loss Bob's detector background counts become the dominant contribution. The lower solid curve (blue) shows QBER under our attack when only Bob's sifted key rate is kept the same as before the attack. The upper solid curve (red) additionally keeps the same sifted key rates conditioned on each polarization sent by Alice, which more closely mimics a realistic system operation (see Ref.~\onlinecite{sajeed2015a} for details).}
  \label{fig:free-space-qber-vs-loss}
\end{figure}

{\bf \noindent Predicted attack on free-space QKD system with damaged pinhole.} We model a practical faked-state attack as described in Ref.~\onlinecite{sajeed2015a}. We assume a part of Eve is situated outside Alice and measures the quantum states coming out. Then, another part of her regenerates the measured quantum states as attenuated coherent pulses and sends them to Bob, tilting her beam at an angle such that it has a relatively higher probability of being detected by the desired detector. Eve has information about Bob's receiver characteristics after the laser damage, and only uses devices available in today's technology \cite{sajeed2015a}. For example, let's assume Eve sends a horizontally polarized light pulse. In this case, she should choose her tilt angle $(\phi,\theta)$ from a subset $\tilde{H}$ selected in such a way that the efficiency $\eta_h(\tilde{H})$ of Bob's horizontal channel in $\tilde{H}$ is as high as possible, in order to maximize mutual information Eve--Bob. On the other hand, if Bob measures in the opposite (DA) basis, the detection probabilities in the D and A channels $\eta_d(\tilde{H})$ and $\eta_a(\tilde{H})$ should be as low as possible, to minimize QBER. Thus, to find attack angles for the horizontally polarized light, we choose $\tilde{H}$ that satisfies $\eta_h(\tilde{H}) \geq 0.6$ and $\delta(\tilde{H}) = \min \left\{ \frac{\eta_h(\tilde{H})}{\eta_{d}(\tilde{H})}, \frac{\eta_h(\tilde{H})}{\eta_{a}(\tilde{H})} \right\} \geq 100$. Similarly, for V, D and A polarized pulses, we choose attack angles that satisfy $\eta_v(\tilde{V}) \geq 0.03$, $\delta(\tilde{V}) \geq 4.5$; $\eta_d(\tilde{D}) \geq 0.6$, $\delta(\tilde{D}) \geq 120$; $\eta_a(\tilde{A}) \geq 0.2$, $\delta(\tilde{A}) \geq 22$. These subsets of angles are shown in the rightmost plot in \cref{fig:free-space-mismatch-plots}b. Note that the thresholds $\eta$ and $\delta$ used here are not optimal and have been picked manually. However, they satisfy the required conditions to successfully perform the faked-state attack with a  resultant QBER $\leq 6.6\%$ in $1$--$15~\deci\bel$ channel loss range, as shown in \cref{fig:free-space-qber-vs-loss}. In the simulation, we assumed that Alice--Bob and Alice--Eve fidelity $F = 0.9831$ \cite{bourgoin2015,sajeed2015a}, while Eve--Bob experimentally measured $F = 0.9904$. All other assumptions were the same as in Ref.~\onlinecite{sajeed2015a}.

\section{Deviation that has led to a denial-of-service, and how we avoided it}
\label{sec:damage-control-fiber-optic}

\noindent Alice's setup in Clavis2 (\cref{fig:fiber-optic-setup}a) consists of more than 20 discrete optical components: over 10 fiber connectors, 3 beamsplitters, 3 detectors, 2 variable attenuators, a bandpass filter, phase modulator, and Faraday mirror. As discussed in main text, certain deviations in any of these components lead to the denial-of-service, while other deviations result in opening different security loopholes. Attacker's goal, and the goal of a thorough security tester, is to do one's best to avoid the former and demonstrate the latter. Parameters of the damaging laser illumination can and should be varied to reach this goal.

When we began testing the system components for laser damage, the synchronization detector D$_\text{sync}$ initially presented an obstacle. This detector was based on an optical receiver module (Fujitsu FRM5W232BS) incorporating an avalanche photodiode biased below breakdown at $>\! 30~\volt$, providing an avalanche multiplication factor $\approx\! 6$. It only took about $6~\milli\watt$ of optical power at the photodiode (translating to about $0.15~\watt$ at Alice's entrance) to die. It stopped providing the synchronization signal for Alice and thus broke the system, i.e.,\ led to the denial-of-service. After an investigation, it turned out that the energy that killed it was chiefly provided by its high-voltage electrical bias circuit and not the optical signal. The bias circuit was based on a specialised integrated circuit with overcurrent protection (Maxim Integrated MAX1932ETC) followed by an LC low-pass filter with inductor $\text{L} = 330~\micro\henry$ and capacitor $\text{C} = 0.47~\micro\farad$. If the optical power is applied suddenly, with sub-nanosecond rise time, it momentarily induces a large photocurrent supplied from C that destroys the avalanche photodiode. If, however, the optical power is applied gradually, with millisecond rise time, C discharges slowly and then the relatively slow overcurrent protection reacts in the integrated circuit, lowers the bias voltage and saves the photodiode. We thus added a manual shutter to the EDFA to make the damaging power rise from zero slowly, allowing D$_\text{sync}$ to easily withstand the optical power used in our attack while being electrically powered up. Another solution could be to damage the system when it is without electrical power. It can also be said that we could choose to selectively damage one of two components in Alice, albeit one of them bricking the system.

We ran our damage tests with VOA2 (OZ~Optics DD-600-11-1300/1550-9/125-S-40-3S3S-1-1-485:1-5-MC/IIC) set at $2~\deci\bel$, because this is what the manufacturer's QKD software available for the research system Clavis2 set it at. The support of the pulse-energy-monitoring countermeasure was not implemented in this software \cite{sajeed2015}. In contrast, the manufacturer's factory-calibration software supported it fully and set VOA2 between $2$ and $\approx\! 15~\deci\bel$, complementary to the channel loss, in order to maintain constant power at the three Alice's detectors D$_\text{pulse}$, D$_\text{sync}$, and D$_\text{cw}$. The higher settings of VOA2 would require more laser power to damage D$_\text{pulse}$. However, D$_\text{pulse}$ could also be damaged during the system start-up time, when it sends the homing command to VOA2. The homing command causes it to traverse its lowest attenuation values for a few seconds, likely being sufficient for Eve to do the damage at already demonstrated power levels.

\section{Real-time video recording of laser damage to the spatial filter inside Bob's setup}
\label{sec:video-pinhole-damage}

\noindent Download the video at \url{http://vad1.com/pinhole-laser-damage-20140825.wmv} (Windows Media Video, 14.4~MiB) or \url{http://vad1.com/pinhole-laser-damage-20140825.ppsx} (PowerPoint Show, 17.0~MiB). The video shows the spatial filter (Thorlabs P20S) illuminated by $3.6~\watt$ c.w.\ $810~\nano\meter$ laser beam for $10~\second$, focused in a spot much wider than the original pinhole diameter of $20~\micro\meter$. This is a filter sample with a slightly smaller original pinhole diameter than the one used to obtain efficiency mismatch data in this article and shown in \cref{fig:free-space-setup}b. The samples were otherwise of the same type and damaged under the same conditions. The video was taken via a mirror lowered inside Bob's setup. The pinhole plane was imaged from the front side at an angle slightly off normal, in order for the mirror not to obstruct the damaging beam. Canon MP-E $65~\milli\meter$ lens was used at $2.8\times$ magnification and f/16 lens aperture (f/60 effective aperture), with Canon EOS~7D camera body. The pinhole was brightly lit sideways with a fiber-optic illuminator bundle, in order to bring up detail. During the laser exposure, the steel foil can be seen deforming from heat, popping out of focus and apparently shifting laterally in the image; however the lateral shift is an artefact of the camera's angle of view being off-normal. After the laser is switched off, the foil cools and returns to the original position, now with about $150~\micro\meter$ diameter hole in it. Sound was added later for an artistic effect. 

\def\bibsection{\medskip\begin{center}\rule{0.5\columnwidth}{.8pt}\end{center}\medskip} %fixes RevTeX bug with last-page layout, by redefining bibliography separator.
%\bibliography{bibtex_library}


\end{document}